\documentclass[amsmath,prl,amssymb,floatfix,twocolumn]{revtex4}
\usepackage{amsthm,amsfonts,graphicx,verbatim,epsfig}
\usepackage{amssymb}

\newcommand{\D}{{\rm d}}

\newcommand{\dd}{{\rm d}}

\newcommand{\ra}{\rangle}
\newcommand{\la}{\langle}

\begin{document}

\title{Novel Mechanisms for repulsive Casimir forces}
\author{Oded Kenneth$^{1}$ and Israel Klich$^{2}$}%
\affiliation{(1) Department of Physics, Technion, Haifa 32000 Israel \\
(2) Department of Physics, University of Virginia,
Charlottesville, 22904,VA}

\begin{abstract}
We present two novel models for repulsive Casimir interaction between positive perturbations. One example
relies on non locality of the dielectric response and one relies
on interference between (attractive) modes. Such examples are
impossible to achieve in 1d massless theories, as they are prohibited by a
generalization of the result of Lambrecht, Jaekel and Reynaid \cite{Lambrecht Jaekel Reynaud 97} to
non local dielectrics and to negative perturbations.
\end{abstract}
\maketitle \vskip 2mm

% Keywords: {\it Casimir , Quantum Fluctuations , Van der Waals ,
%Zero-Point Energy}
          %\tableofcontents

%Separable potentials arise in

\section{Introduction}

One of the intriguing aspects of the Casimir effect is it's subtle
dependence on geometry and type of boundaries. A lot of effort and
progress has been made recently
 in quantifying the force in various situations, mostly with
experimental applications in mind utilizing scattering
approximations as well as numerical ones. From both practical and
theoretical point of view, reversing the sign of the force from
attractive to repulsive is of particular interest. This
requires considering non-symmetric situations to avoid the
theorem \cite{KennethKlich06,Bachas} ensuring attraction between
opposing bodies having the same properties. The dependence of
the sign on asymmetric boundary conditions was studied in many
works, most recently, in the context of critical Casimir forces
\cite{Schmidt Diehl} where the direction of the force for various
boundary conditions was explored for interacting theories under a
renormalization group flow.

 The
intuition regarding the sign of Casimir forces may be most easily
understood from second order perturbation theory. This includes
pair-wise summation for dielectric bodies, or the two scattering
approximation of \cite{BalianDuplantier78}. The result of this
approach is attraction for any two dielectrics in vacuum. On the
other hand, perfectly magnetically polarizable and perfectly
electrically polarizable atoms will repel each other
\cite{Feinberg}, with the consequence of repulsion between a
perfectly conducting and perfectly permeable materials
\cite{Boyer74,Alves,daSilva}.

For many purposes it is convenient to "soften" the boundary
conditions, by considering the interaction of the free field with
the material, as a perturbation on the free field wave equation.
Consider for example, two dielectrics $A$ and $B$
immersed in a medium $M$, of dielectric constant $\epsilon_M$.
The wave equation in this case is written as
\begin{eqnarray}\label{wa}
\nabla\times\nabla\times A+\omega^2 \epsilon(x)A=0
\end{eqnarray}
where $\omega^2\epsilon(x)=\omega^2\epsilon_M+V_A+V_B$
is a sum of the constant background term $\omega^2\epsilon_M$,
and two perturbation terms
$V_A(x)=\omega^2\epsilon_A(x)-\omega^2\epsilon_M,
 V_B(x)=\omega^2\epsilon_B(x)-\omega^2\epsilon_M$
representing the response of the two bodies.

Second order perturbation theory for the energy of the system
gives (ignoring self energy terms which do not affect the mutual force)
\begin{eqnarray}\label{de1}
E^{(2)}=-\int{d\omega\over4\pi}Tr(V_AG_0V_BG_0)
\end{eqnarray}
More explicitly this may be written as the integral (see e.g.
\cite{Golestanian,Kenneth Klich 07,Milton}) as
\begin{eqnarray}\label{de2}&
  E^{(2)}\propto \\ \nonumber & -\int{\D\omega\over4\pi}\int_A\D x\int_B\D y
  V_A(x,i\omega)V_B(y,i\omega)\sum_{ij}(G_{0ij}(x-y))^2
\end{eqnarray}
where ${G_0}({\bf x},{\bf x}')_{ij}=\langle {\bf x}|{1\over
\nabla\times\nabla\times+\omega^2\epsilon_M}|{\bf x}'\rangle_{ij}$
is the Euclidian (Wick rotated) propagator in the medium $M$.
Since ${G_0}({\bf x},{\bf x}')_{ij}$ is real, Eq. (\ref{de2})
shows that the sign of the interaction energy depends only on the
signs of the perturbations $V_A,V_B$ as
$sgn[E^{(2)}]=-sgn[V_A]sgn[V_B]$. 
%%
%%{\bf This only shows sign of energy, one needs to also show monotoncity}
%%
%% A more sophisticated argument based on eq(\ref{de1}) and on the fact
%% that $G_0$ is a positive operator allows generalizing
%% the last conclusion to the case where the perturbations $V_A,V_B$
%% are given by arbitrary operetors of definite sign.(An hermitian operator $V$
%% is said to be positive/negative iff all its nonzero eigenvalues are positive/negative.
%% This is equivalent to requiring $\la\psi|V|\psi\ra\geq 0$ for all states $|\psi\ra$.)
%%
Since $E^{(2)}$ vanish for infinitely separated bodies,
an immediate conclusion is that at least at large distances,
perturbations of similar signs attract while perturbations of
opposite sign repel. It seems natural to conjecture this to
hold also at arbitrary distances. This however does not follow from
the above argument, since the energy need not be a monotonic function
of the distance.

Eq. \eqref{de2} is an easy way of understanding the well known
situation where the Hamaker constant is negative, when considering
materials $A,B$ such that $\epsilon_A<\epsilon_M<\epsilon_B$.

%% One can easily check that $(G_{0ij}(x-y))^2$ decays with distance, and
%% therefore, if $V_A,V_B$ have the same sign, bodies would like to
%% minimize the distances between them to minimize energy.

Using the more abstract expression Eq. (\ref{de1}) and the fact
that $G_0$ is a positive operator allows generalizing the above
arguments to the case where the perturbations $V_A,V_B$ are given
by arbitrary operators of definite sign.(An hermitian operator $V$
is said to be positive/negative iff all its nonzero eigenvalues
are positive/negative. This is equivalent to requiring
$\la\psi|V|\psi\ra> 0$ for all states $|\psi\ra$.)

One example of such generalized perturbation is to assume nonlocal
dielectric functions. As another example imagine the bodies $A,B$
to have nontrivial magnetic permeability. In this case Eq.
(\ref{wa}) would have to include the extra term
$\nabla\times({1\over\mu(x)}-1)\nabla\times A$ which may be
interpreted as a perturbation $VA$ with $V$ a differential
operator. For $\mu>1$ one may show that this differential operator
is negative. Thus the result \cite{Feinberg,Boyer74,Alves,daSilva}.
that a dielectric wall repel from a permeable one is consistent 
with the mnemonic that "positive" perturbations attract one another 
while positive perturbation will repel a negative one.

{\bf }

All known examples of repulsive Casimir forces can be traced to
such a mechanism.
One may thus expect that the sign of the
Casimir force is the product of the signs of the perturbations,
whenever these perturbations have a definite sign.

There are two general statements regarding the sign of the force
at all distances: In 1d, Lambrecht, Jaekel and Reynaud
\cite{Lambrecht Jaekel Reynaud 97} have shown that dielectrics
attract under very general conditions. We shall show below that
this statement also hold for general positive perturbations in 1d.
Unfortunately, this argument does not extend to higher dimensions.
The only available result to date is that under conditions of
mirror image symmetry, the force is always attractive
\cite{KennethKlich06,Bachas}.

%% In the attraction theorem, the precise nature of perturbations
%% does not play a role, and they can be any general operators, such
%% as non-local dielectric functions. Thus, it is tempting to expect,
%% in addition, that positive perturbations attract each other, where
%% "positive" perturbation $V$ is define by the condition that
%% $\la\psi|V|\psi\ra> 0$. One may thus expect that the sign of the
%% Casimir force is the product of the signs of the perturbations,
%% whenever these perturbations have a definite sign.
%%
%% Here we emphasize that the statements of \cite{Lambrecht Jaekel
%% Reynaud 97} also hold for general positive perturbations.

A natural question that rises is: is attraction always guaranteed
between a pair of positive perturbations or a pair of negative
perturbations?  If not, what other mechanism can be responsible for repulsion ?
%and can we find necessary conditions for attraction to hold?

In this paper we give a partial answer to the question raised
above: We show that this intuition is, in fact, correct in 1D,
i.e. any two perturbations (in the general sense) having the same
sign attract, by which we generalize the result \cite{Lambrecht
Jaekel Reynaud 97, Lambrecht}. On the other hand, the situation in
higher dimensions is not as simple: In fact, we find examples of
positive perturbations which yield repulsion. These examples, do
not rely on the relative sign of the perturbation as previous
examples do, and thus provide new mechanisms for repulsion. While
the examples we present are "toy" models, it is important to
stress that both examples may be thought of as extreme limits of
physical perturbations.

A convenient starting point for our analysis is the TGTG
representation for the energy \cite{KennethKlich06}: We consider
the interaction between two bodies $A$ and $B$ which are defined
through their dielectric susceptibilities
$\chi_A=\epsilon_A-1,\chi_B=\epsilon_B-1$ respectively or any
generalized perturbation (i.e. $\chi_A$ can be an differential or integral operator). The regularized Casimir energy is given
by:
\begin{eqnarray}\label{TGTG}&
E_C(a)=\int_0^{\infty}{\dd \omega\over 2\pi} \log\det(1-T_A{G_0}
T_B{G_0} ).
\end{eqnarray}
Here $G_0$ is the free (Helmholtz) Green's function, given by
${G_0}({\bf x},{\bf x}')=\langle {\bf x}|{1\over
-\nabla^2+\omega^2}|{\bf x}'\rangle$, for a scalar field (or
${G_0}({\bf x},{\bf x}')_{ij}=\langle {\bf x}|{1\over
\nabla\times\nabla\times+\omega^2}|{\bf x}'\rangle_{ij}$ when
considering the electromagnetic field). The operators
$T_{A},T_{B}$ are a Wick rotated version of the the transition
operators appearing in the Lippmann-Schwinger equation. They are
given by
$$T_{i}={\omega^2\over 1+\omega^2\chi_{i} {G_0} }\chi_{i},\,\,\,\,\,\,i=A,B.$$
\begin{comment}
 relate the green's
function in presence of the dielectric with the free one via
\begin{eqnarray}\label{G=G_0+G_0TG_0}
G(\omega)=G_0(\omega)+G_0(\omega)T(\omega)G_0(\omega)
\end{eqnarray}

Up to Wick rotation the operators $T_{A},T_{B}$ are the transition
operators appearing in the Lippmann-Schwinger equation:
\begin{eqnarray}
  \psi_k=\phi_k+\int\D^3 k G_0(k)T(k,k')\phi_{k'}
\end{eqnarray}
where $\phi_k$ are freely propagating waves, while $\psi_k$ are
the solutions in presence of the scatterer.

The Wick rotation $T(\omega)\rightarrow T(i\omega)$ has the effect
of turning $T$ into hermitian operators as well as of avoiding
potential singularities which may occur at real frequencies. Note
that the matrix elements $\langle x|T_A|x'\rangle$ and $\langle
x|T_B|x'\rangle$ of the operators $T_A,T_B$ are supported on the
volumes of bodies $A,B$ respectively (i.e. vanish when $x,x'\notin
A$ for $T_A$ for example). Therefore, when the propagator $G_0$ is
inserted between them as in $T_AG_0T_B$, it is enough to consider
$G_0(x_a,x_b)$ with $x_a\in A,x_b\in B$.
\end{comment}
In general, the determinant formula for the Casimir energy is hard
to analyze because it involves the determinant of an infinite
dimensional operator. However, the fact that the TGTG determinant
in \eqref{TGTG} is well defined \cite{KennethKlich06}, means that
in practice only a finite dimensional subspace gives significant
contribution to the force. Such a finite dimensional reduction
allows arbitrary good numerical evaluation of the force. This suggests that the question of
the possible sign of the force may also be addressed within the
context of finite dimensional operators. In any case, to get the
correct sign of the force, one should be careful when using infinite
dimensional tricks such as the "identity" $\sum n=-{1\over 12}<0$.

In this paper we concentrate on two situations where the determinant in
\eqref{TGTG} reduces to a strictly finite dimensional one. The
first is the case of the so called separable potentials. These are
(typically non-local) potentials which by construction are given
by finite rank operators and therefore interact only with a finite
dimensional subspace of states. The second is the case of one
dimensional systems. At each given energy $\omega$ a one
dimensional field $\phi$ has only two modes: the left and the
right mover state  $|L\ra, |R\ra$. (A multicomponent
$\phi=(\phi_1,\phi_2,...\phi_n)$ would have $2n$ modes). The
determinant \eqref{TGTG} then becomes an ordinary finite
dimensional one.

Below, we show that both of these models
allow simple examples where the interaction
%%force may be computed exactly and
%%may yield direction opposite to the naive expectation.
%%
energy between a pair of local positive perturbations may be computed exactly. We then show how the perturbations can be chosen in a way to yield a non-monotonic energy dependence on distance,
implying that the force can reverse its direction, even when both constraints are positive.

\section{Attraction in 1D for general positive perturbations} The purpose of this section is to establish that in 1D,
any two positive perturbations attract each other. In one
dimensional systems the TGTG formula for the Casimir interaction
between objects $A,B$ situated a distance $a$ apart reduces to
well known relation \cite{Kats}:
\begin{eqnarray}
E=\int{d\omega\over2\pi}\log\left(
1-e^{-2a\omega}\tilde{r}_A(i\omega) r_B(i\omega)\right),
\end{eqnarray}
where $r_A(i\omega),r_B(i\omega)$ are reflection coefficients for plane waves scattering on the bodies $A,B$, evaluated on the imaginary frequency axis. 
It was shown in \cite{Lambrecht Jaekel Reynaud 97} that for a material whose dielectric function is local, 
$r(i\omega)<0,\forall\omega\in\mathbb{R}$, which has the immediate
consequence that in one dimension any two local dielectrics attract.
Here we show that this result in fact applies to any kind of
positive perturbation, including the important case of non-local
dielectric functions.

To do so, we first write the reflection coefficient from a 1d potential as
the matrix element
\begin{eqnarray}
r(\omega)={-i\over2\omega}\la R|T|L\ra={-i\over2\omega}\int dx dx' (e^{-i\omega
x})^*T(x,x',\omega)e^{i\omega x'}
\end{eqnarray}
Thus after Wick rotation we have $r(i\omega)={-1\over2\omega}\int
dx dx' e^{-\omega x}T(x,x',i\omega)e^{-\omega x'}$. (The integral
converge since we consider only potentials of compact support and
hence also $T$ is of compact support.).
It has been shown
\cite{Kenneth Klich 07}, that a positive potential $V>0$, implies
that at imaginary frequencies the Lippmann-Schwinger operator is
positive $T(i\omega)>0$ 
{
\footnote{This follows from writing $T=\sqrt{V}{1\over 1+\sqrt{V}G_0\sqrt{V}}\sqrt{V}$ for $V>0$.}} .
Note that $r(i\omega)$ is of the general form 
$-\langle\psi|T(i\omega)|\psi\rangle$ so that the positivity
of $T(i\omega)$ implies that $r(i\omega)<0$.

Similar arguments may be applied to a 1d multi-component field
$\phi=(\phi_1,\phi_2,...\phi_n)$, assuming all components are massless
(or have exactly the same mass). The reflection coefficients
$r_{A,B}$ in this case turn into $n\times n$ matrices
which by the argument given above are strictly negative matrices.
The energy may be expressed as
\begin{eqnarray}
E=\int{d\omega\over2\pi}\log\det\left(1-e^{-2a\omega}\tilde{r}_A(i\omega)
r_B(i\omega)\right),
\end{eqnarray}
where the determinant is an ordinary $n\times n$ one. Since  ${\rm
spec}(r_Ar_B)={\rm spec}(\sqrt{-r_A}(-r_B)\sqrt{-r_A})=\{
\lambda_1,\lambda_2,..\lambda_n\}$ is positive,the energy can be
written as
\begin{eqnarray}
E=\int{d\omega\over2\pi}\sum_{k=0}^n\log\left(1-e^{-2a\omega}\lambda_k(\omega)\right)
\end{eqnarray}
where $0\leq \lambda_1,\lambda_2,..\lambda_n<1$ are
$a$-independent. 
(The inequality  $0\leq\lambda<1$ has been shown in \cite{Kenneth Klich 07}) 
We immediately see that:
\begin{eqnarray}\label{force 1d}
F_C=-\partial_a E_C=-\int{d\omega\over2\pi}\sum_{k=0}^n{2\omega
e^{-2a\omega}\lambda_k(\omega)\over
\left(1-e^{-2a\omega}\lambda_k(\omega)\right)}<0
\end{eqnarray}

%%
%% Since, in this case, again $G_0$ acts as a scalar, we have attraction
%% between a pair of positive perturbations.
%%

Alternatively, one could arrive at the same conclusion directly
from (\ref{TGTG}) by noting that in the 1d scattering basis $G_0$
appearing in the expression $TGTG$ is essentially the c-number
$e^{-a\omega}$. %%
%% Indeed, computing $G_0$ and the $T$s in the
%% scattering wave basis, shows that $G_0$ appearing in the
%% expression $TGTG$ is essentially the c-number $e^{-a\omega}$.
%%
Since positivity of $V_A,V_B$ guarantee that $T_A,T_B$ are also
positive operators, as shown in \cite{KennethKlich06} it is enough
to have required property of $TGTG=e^{-2ka}T_AT_B$. However the
${\rm spec}(T_AT_B)={\rm spec}(\sqrt{T_A}r_B\sqrt{T_A})$ is
positive, therefore the energy can be written as
\begin{eqnarray}
E=\int{d\omega\over2\pi}\sum_{k=0}^n\log\left(1-e^{-2a\omega}\lambda_k(\omega)\right)
\end{eqnarray}
where $0\leq \lambda_1,\lambda_2,..\lambda_n<1$ are
$a$-independent.

\section{Interference induced Casimir repulsion for fields with different dispersion relations}

Formally, a field in a space of arbitrary dimension may be thought
of as a field in one dimension by considering the transverse
momentum $k_\perp$ as if it was an internal continuous index. Thus
one may ask why doesn't the 1d result \eqref{force 1d} extend to
higher dimensions.

The answer has nothing to do with the fact that there are
infinitely many $k_\perp$ modes. Rather, it is related to the fact
that in strict 1d we have (implicitly) assumed that in free space
(i.e. outside the scatterers) all the component of
$\phi=(\phi_1,\phi_2,...\phi_n)$ share the same (zero) potential.
This is in contrast to the higher dimensional problem where the
term $k_\perp^2$ in the lagrangian serves to distinguish between
the different scattering channels even in empty space and makes
the fields in the 1d picture have different masses. As a result,
 the ``1d momentum'' $k_\|$ is no longer conserved,
which from the 1d point of view is a source of great extra
complications.

\begin{figure}[tbh]
\includegraphics[scale=0.4]{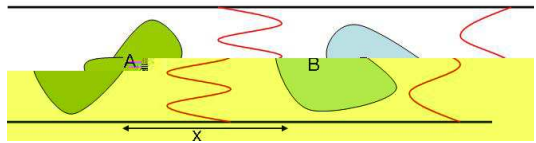}%{Bodies_in_Cylinder.eps}
\caption{Bodies $A$ and $B$ in a confined space, transverse modes
are quantized. In this situation, only low transverse modes
contribute. The problem becomes equivalent to a finite number of
1d fields with different masses.}\label{fig1a}
\end{figure}

A simple toy model corresponding to the above picture may be
constructed by taking $\phi=(\phi_1,\phi_2,...\phi_n)$ to be 1d
fields of masses $m_1,m_2,...m_n$. Let us denote
$k_i=\sqrt{\omega^2+m_i^2}$ and let $K$ be the diagonal matrix
having $k_i$ as its eigenvalues. The (imaginary frequency) free
propagator may then be expressed as $G_0={1\over2K}e^{-K|x-x'|}$.

Consider enforcing on $\phi$ the Dirichlet type boundary
conditions: $\sum \alpha_i\phi_i|_{x=a}=0,\;\; \sum
\beta_i\phi_i|_{x=b}=0$. Here
$\alpha=(\alpha_1,\alpha_2,....\alpha_n)$ and
$\beta=(\beta_1,\beta_2,....\beta_n)$ are some constant vectors.
This may be obtained by assuming the (positive) potential
$V=V_A+V_B$
$$V_A=\lambda\alpha\otimes\alpha^\dag \delta(x-x_a),\;
V_B=\lambda\beta\otimes\beta^\dag \delta(x-x_b)$$
with $\lambda\rightarrow\infty$.
One then finds the reflection coefficients from the two potentials:
\begin{eqnarray}&
r_A(i\omega)=\tilde{r}_A(i\omega)=
-{\tilde{\alpha}\otimes\tilde{\alpha}^\dag\over
|\tilde{\alpha}|^2+2\omega/\lambda},\\ \nonumber &
r_B(i\omega)=\tilde{r}_B(i\omega)=
-{\tilde{\beta}\otimes\tilde{\beta}^\dag\over|\tilde{\beta}|^2+2\omega/\lambda}
\end{eqnarray}
where we used the notation $\tilde{\alpha}=\sqrt{\omega\over K}\alpha=
(\alpha_1\sqrt{\omega\over k_1},\alpha_2\sqrt{\omega\over k_2},...,
\alpha_n\sqrt{\omega\over k_n})$
and similarly $\tilde{\beta}=\sqrt{\omega\over K}\beta$.
{We note, parenthetically,  that a more general dispersion require $\sqrt{\omega\over k}\rightarrow\sqrt{dk\over d\omega}$ 
i.e. the group velocity is the extra factor.}
Calculating the determinant one obtains
\begin{eqnarray}\label{det modes}&
\det(1-r_Ae^{-K|x_a-x_b|}\tilde{r}_Be^{-K|x_a-x_b|})= \\
\nonumber & 1-\left(
{\langle\tilde{\alpha}|e^{-K|x_a-x_b|}|\tilde{\beta}\rangle\over
|\tilde{\alpha}|\cdot|\tilde{\beta}|}\right)^2
\end{eqnarray}

\begin{figure}[tbh]
\includegraphics[scale=0.6]{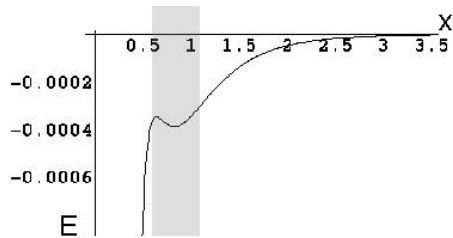}%{rce.eps}
\caption{Casimir energy as a function of distance (measured in
units where $m_1\equiv1$), obtained from integration of Eq. \eqref{det
modes} over frequencies. Grey area marks the local energy
minimum.}\label{g}
\end{figure}

It is quite easy to see that this expression need not be monotonic
as a function of the distance $x=|x_a-x_b|$. To check the behavior
of the total ($\omega$-integrated) energy is more complicated.
However by choosing special values for the parameters one may
check numerically that the resultant energy need not be monotonic.
The following graph, Fig. \ref{g}, shows the interaction energy
for the case
 $n=2$ with $m_2/m_1=\alpha_2/\alpha_1=-(\beta_2/\beta_1)=5$ .

We stress that each of the field modes $\phi_1$ and $\phi_2$ on
each own would produce an attractive force if the other was
somehow turned off (e.g. by giving it very large mass). The
repulsion seen in Fig. \ref{g} may therefore be understood as an
{\bf interference} between the modes (or, more precisely, from the
dependence of the interference term on the separation). As a side
remark, we mention that the mathematics involved here is somewhat
reminiscent to that of the Glashow- Iliopoulous-Maiani (GIM)
mechanism \cite{GIM}, although actually applying it to quark
mediated weak interactions would be, of course, completely
unrealistic.

For a realistic way to obtain such a model, consider e.g. objects
situated inside a circular cylinder of radius $R$ as in Fig.
 \ref{fig1a}. Standard mode expansion of a field leaving in this
cylinder would reduce it to a series of 1d fields of masses
$m_n=\zeta_n/R$ with $\zeta_{n,m}$ zeroes of Bessel functions $J_m$
and of each derivative for TM and TE modes respectively.
Standard boundary conditions on the bodies would demand specific
(shape dependent) combinations of the 1d fields to vanish.

The fact that for very large and for very small distances the
force attracts is a general feature not special to the specific
model.
At very large distances $x\gg1$ only the channels of lowest mass
contribute to the force. Having only this single mass, the above
no-go theorem applies and we obtain attraction.
At very small distances $x\ll1$ the integrand decays very slowly
with $\omega$ so that most of the contribution to the force comes
from very large frequencies $\omega\sim{1\over x}$. At such
frequencies all the $k_i$'s become practically equal:
$k_i=\sqrt{m_i^2+\omega^2}\sim\omega\;\forall i$ and our theorem
again imply attraction.

(E.g. in the above model $E\simeq -{1\over4\pi
x}Polylog[2,{(\alpha\cdot\beta)^2\over \alpha^2 \beta^2}]$.)
At intermediate distances the force may however change sign as the
above example demonstrates.

 It should be remarked however that this reasoning relies on
the assumption of having only finitely many channels. In an
infinite channel system it is conceivable that as $x$ becomes
smaller more and more new channels (of higher $m$) become
relevant, making the above argument false.

The short distance
attraction argument may also become false in case the reflection
coefficients decay very rapidly as $\omega\rightarrow\infty$. Also
we remark that using fine tuning one might be able to construct a
model where the leading short distance attractive term vanishes,
in which case the resultant force can remain repulsive down to
$x=0$. For example: Suppose one constructs a model where the
$\alpha$'s and $\beta$'s depend on $\omega$ in such a way that
$\tilde{\alpha},\tilde{\beta}$ are constants and take ($n=2$)
$\alpha_2/\alpha_1=-\beta_2/\beta_1=1, m_1=0,m_2\neq0$ then the
attractive term which usually dominates at $x\rightarrow0$
vanishes and the next to leading term result in a force which is
attractive down to $x=0$. However any slight change of the
parameters will make the usual leading term re-appear and thus
turn the force into an attractive one (at $x=0$).

\section{Non-locality: Repulsion between Separable Potentials}

Our next repulsive example shows that for $d>1$ non-locality of
dielectric functions, may also, in principle, result in repulsive
behavior. The effect of spatial dispersion on Casimir forces
between metals has been of interest for some time
\cite{Kats,Sernelius}. It is particularly important in metals
where long range density correlations in the medium can be present
and substantially change the scaling of the force
\cite{Dobson06}.

Our example here is very different from the situations considered
in the references above. Our toy model is based on so called
"separable potentials". Such potentials arise in variety of
situations in physics and in mathematics, and were first
introduced for Casimir type problems by Jaffe and Williamson
\cite{Jaffe Williamson}. They correspond to "rank 1"
perturbations, and can be written as $V=|f\ra\la f|$ for some
function $f$, or in $x$ space notation as $V(x,x')=f(x)f^*(x')$.
%% Most importantly %%
Such separable potentials describe the
interactions involved in Feschbach resonances, and play a major
role in the properties of fermi condensates. Note, also, that such $V$ is a positive operator by
construction since $\la\psi|V|\psi\ra=|\la\psi|f\ra|^2>0$ for any
$\psi$.

\begin{figure}
\includegraphics[scale=0.4]{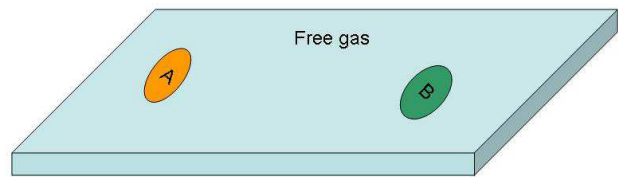}
\caption{An effective interaction with a separable potential can
happen when a free field is perturbed by tunnelling amplitude into
traps, $A$ and $B$ in which the field may have bound
states.}\label{traps}
\end{figure}

The situation we consider is of a scalar field, which can tunnel
into traps where it has a bound state Fig. \ref{traps}. Thus, in
our case the perturbations are traps. Integrating out degrees of
freedom in these traps will
 generate a perturbation of the form $\int\D x\D
 x'\phi(x)t(x)G_{bound}(x,x')t^*(x')\phi(x')$ for the field outside the trap,
 where $G_{bound}$ is the green's function of the trap, and $t(x)$
 is the tunnelling element.

Close to resonance, $G_{bound}$ may be approximated as:
$G_{bound}\sim{f_b(x)f_b^*(x')\over E-E_b}$ where $f_b$ is the
wave function of a bound state, and when $E$ is close to $E_b$.
Thus, the effective interaction with the trap is through the
separable potential $V=f(x)f^*(x')$, where $f(x)=\sqrt{{1\over
|E-E_b|}}t(x)f_b(x)$.

For such potentials, one may readily calculate the $T$ operators,
which turn out to be also of rank one, and so the interaction
energy is easily calculated (One may also look at sums of
separable potentials, corresponding to finite rank perturbations,
which can be computed in the same way).

In case the free field interacts with the perturbation
$V_{bound}\propto |f\ra\la f|$. We have:
\begin{eqnarray}&
T={1\over 1+VG_0 }V=\sqrt{V}{1\over 1+\sqrt{V}G_0\sqrt{V}
}\sqrt{V}=\\ \nonumber &{1\over 1+\la f|G_0|f\ra}|f\ra\la f|
\end{eqnarray}
And so,
\begin{eqnarray}\label{seperable determinant}&
\log\det(1-T_AG_0T_BG_0)=\\ \nonumber & \log(1-{1\over 1+\la
f_A|G_0|f_A\ra}{1\over 1+\la f_B|G_0|f_B\ra}|\la
f_B|G_0|f_A\ra|^2)
\end{eqnarray}
thus, since the (positive) terms ${1\over 1+\la
f_A|G_0|f_A\ra}{1\over 1+\la f_B|G_0|f_B\ra}$ do not depend on the
distance, to find the direction of the force it is enough to
consider $|\la f_B|G_0|f_A\ra|$.

\begin{figure}[tbh]
\includegraphics[scale=0.4]{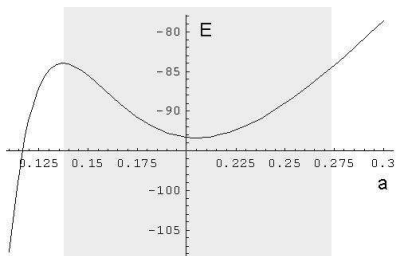}
\caption{Casimir energy as a function of distance $a$ between
separable potentials $V_A=|f_A\ra\la f_A|$ and $V_A=|f_B\ra\la
f_B|$ for $f_A,f_B$ defined in \eqref{fafb}.}\label{sep}
\end{figure}

As an example, we consider the following functions \footnote{To
avoid difficulties associated with $\delta$-function potential in
3d one should actually assume slightly smeared $\delta$-function.
}:
\begin{eqnarray}\label{fafb}&
f_A=(1+{1\over
\omega^2+1})[\alpha_1\delta(x+a)+\alpha_2\delta(x+a+0.1)]
\\ & \nonumber f_B=(1+{1\over
\omega^2+1})[\beta_1\delta(x-a)+\beta_2\delta(x-a-1)].
\end{eqnarray} Taking also:
$\alpha_1=1,\alpha_2=-4,\beta_1=1,\beta_2=-1$ we get a repulsive
regime as shown in Fig. \ref{sep}.

\section{Summary}

To summarize,
%%  we have studied two new mechanisms for Casimir repulsion. 
we gave a general rule for the sign of the Casimir force at large distances
and considered mechanisms allowing the force to reverse its direction
at shorter distance.
We have shown that such behavior is possible when an
interaction is introduced between fields with different dispersion
relations. 
Such a situation may mimic the interaction between
bodies confined in an infinite cylinder which serves to quantize
the transverse modes, rendering a quasi 1d situation. Secondly, we
have shown that non-locality of a dielectric function may, in an
extreme limit, result in repulsive behavior. This type of
non-locality, typical of separable potentials, may arise in
situations where the bodies represent "traps" into which the field
can tunnel, and interact with a bound state. 

{\bf Acknowledgments:} We would like to thank the KITP for
 hospitality
during part of this work, as well as partial support from the
NSF under Grant No. PHY05-51164. We also thank R. L. Jaffe for
discussions and useful comments regarding separable potentials.

\end{document}